\documentstyle[epsfig,longtable,lscape,pennames,amstex]{aipproc}

\begin{document}

%
% ------------------------- definitions ------------------------
%         

\def\SELEX{\textsc{Selex} }

\def\r2{\langle r^2 \rangle}
\def\Q2{$Q^2$}

\def\gevc{GeV/$c$ }
\def\gev2c2{GeV$^2$/$c^2$}
\def\fm2{\text{fm}^2}

\def\S1385{\ifmmode{ \Sigma(1385)^-}
           \else$\Sigma(1385)^-$\fi}

\def\Journal#1#2#3#4{{#1} {\bf #2}, #3 (#4)}
\def\EPJA{{\em Eur. Phys. J.} A}
\def\EPJC{{\em Eur. Phys. J.} C}
\def\NCA{\em Nuovo Cimento}
\def\NPA{{\em Nucl. Phys.} A}
\def\NPB{{\em Nucl. Phys.} B}
\def\PLA{{\em Phys. Lett.} A}
\def\PLB{{\em Phys. Lett.} B}
\def\PLD{{\em Phys. Lett.} D}
\def\PL{{\em Phys. Lett.}}
\def\PRL{\em Phys. Rev. Lett.}
\def\PREV{\em Phys. Rev.}
\def\PREP{\em Phys. Rep.}
\def\PRD{{\em Phys. Rev.} D}
\def\PRC{{\em Phys. Rev.} C}
\def\ZPC{{\em Z. Phys.} C}
\def\ZPA{{\em Z. Phys.} A}

%
% ----------------------------- paper ---------------------------
%         
         
\title{Hyperon Physics Results from SELEX\footnote{Presented at the
    Workshop on Heavy Quarks at Fixed Target, Fermilab, Oct. 10-12, 1998}}

%\author{Ivo Eschrich}
%\address{Max Planck Institute for Nuclear Physics, Heidelberg, Germany}
%\author{on behalf of the \SELEX Collaboration%} %%\footnotemark}

%-----------------------------------------------------------------------------
%
%
%       SELEX collaboration list    ---   plain-block version
%
%       Ivo Eschrich, MPI Heidelberg
%
%     Last modified : 11/26/98 13:06
%
%
%-----------------------------------------------------------------------------

\author{\large The \textsc{Selex} Collaboration\\}
\author{\small\noindent
  I.~Eschrich$^{i,}$\footnote{\emph{Now at Imperial College, London SW7 2BZ,
    U.K.}},
  N.~Akchurin$^{q}$,
  V.~A.~Andreev$^{k}$,
  A.G.~Atamantchouk$^{k}$,
  M.~Aykac$^{q}$,
  M.Y.~Balatz$^{h}$,
  N.F.~Bondar$^{k}$,
  A.~Bravar$^{v}$,
  M.~Chensheng$^{g}$,
  P.S.~Cooper$^{e}$
% ...
\\
}

\address{\noindent\footnotesize
% ...
$^{e}$ Fermilab, Batavia, IL 60510, U.S.A.\\
% ...
$^{g}$ Institute of High Energy Physics, Beijing, PR China\\
$^{h}$ Institute of Theoretical and Experimental Physics, Moscow, Russia\\
$^{i}$ Max-Planck-Institut f\"ur Kernphysik, 69117 Heidelberg, Germany\\
% ...
$^{k}$ Petersburg Nuclear Physics Institute, St. Petersburg, Russia\\
% ...
$^{q}$ University of Iowa,  Iowa City, Iowa  52242, U.S.A.\\
% ...
$^{v}$ University of Trieste and INFN, Trieste, Italy\\
}

%\lefthead{LEFT head}
%\righthead{RIGHT head}
\maketitle

%
% ------------------------ abstract ---------------------------
%

\begin{abstract}
In parallel to charm hadroproduction the experiment \SELEX (E781)
at Fermilab is pursuing a rich hyperon physics program.
\SELEX employs a 600 GeV/c beam consisting of 50~\% \PgSm\ and \Pgpm\ each.
The three-stage magnetic spectrometer covering $0.1 \le x_F \le 1.0$
features a high-precision silicon vertex system,
broad-coverage particle identification
using TRD and RICH, and a three-stage lead glass photon calorimeter.
First results for the \PgSm\ charge radius, total \PgSm- nucleon
cross sections, and a new upper limit for the radiative width of
the \S1385\ are presented. 
\end{abstract}

%\footnotetext{
%  Ball State University,
%  Bogazici University Istanbul,
%  Carnegie Mellon University,
%  Centro Brasileiro de Pesquisas Fisicas Rio de Janeiro,
%  Fermilab,
%  IHEP Bejing,
%  IHEP Protvino,
%  ITEP Moscow,
%  Max-Planck-Institut f\"ur Kernphysik Heidelberg,
%  Moscow State University,
%  Petersburg Nuclear Physics Institute,
%  Tel Aviv University,
%  Universidad Autonoma de San Luis Potosi,
%  Universidade Federal da Paraiba,
%  University of Bristol,
%  University of Hawaii,
%  University of Iowa,
%  University of Michigan-Flint,
%  University of Rochester,
%  University of Rome 'La Sapienza' and INFN Rome,
%  University of S\~ao Paulo,
%  University of Trieste and INFN Trieste.
%  }

%
% ------------------------ introduction ---------------------------
%
  
\section{Introduction}
\label{sec:hypphys}

In parallel to charm hadroproduction the hyperon beam experiment \SELEX (E781)
at Fermilab is pursuing a rich hyperon physics program. One year after
the end of the 1997 fixed target run, first results are available.

The apparatus
has been described already elsewhere in these proceedings \cite{sasha}.
Ongoing projects include the measurement of hyperon charge radii by
hyperon-electron elastic scattering, the \PgSm-proton total cross section,
Primakoff production of hyperon resonances, \PgSp\ production polarization,
and the search for exotic particles produced by hyperon-induced reactions
to name a few examples. The hyperon physics program is complemented by
analysis of the data taken with \Pgpm\ and proton beams, which has also yielded
first results \cite{a2_ichep}.

%
% ------------------------ Sigma- charge radius ------------------------
%

\section{Charge radii}
\label{sec:radii}

Hadrons as we understand them today are composite systems which we
characterize by their static properties.
One static property which reflects the
phenomenon unique to hadrons -- quark confinement -- is the size
of the particle.

%The definition of the radius of a hadron depends on the probe used to
%measure it. A strong-interaction mean squared radius can be extracted
%from hadron-proton
%collisions. The hadron-electron interaction, on the other hand, yields
%the mean squared charge radius:
Elastic scattering of an electron off a charged hadron is modified from
a point interaction by the form factor $F(Q^2)$ where \Q2 is the
four-momentum transfer squared.
At zero momentum transfer the mean squared charge radius is related to
the slope of the form factor by
\begin{equation*}
  \label{equ:rad_from_ff}
  \r2 = -6\hbar^2 \frac{dF(Q^2)}{dQ^2}\Bigg|_{Q^2=0}.
\end{equation*}
Charge radii are known only for five different hadrons so far.
%(\Pp, \Pn, \Pgpm, \PKm, and \PKz).
The fact that
the \PKm\ radius has been found to be smaller than that of the \Pgpm\
by $\sim 0.1\: \fm2$
suggests that the size of a hadron is related to the flavor composition
of its constituent quarks. There is supporting evidence from a 
study of strong interaction radii \cite{povh90} which finds that
replacing an \textit{up} or \textit{down} quark by a
\textit{strange} quark in a baryon decreases its radius by approximately
0.08~fm$^2$.
Consequently the \PgSm\ radius should be smaller than
the proton radius, and larger than the \PgXm.
The definition of a strong-interaction radius, however, is model-dependent.
The significance of the above observation is therefore limited unless
validated by a systematic study of hyperon charge radii.

For hadron-electron elastic scattering,
two hits in the negative and none in the positive half of a hodoscope
downstream of the second magnet in coincidence with a multiplicity of
two in a set of scintillation counters 3~cm downstream of the target
constituted a valid trigger condition.
The typical trigger rate at this level was 3000 per 20-second spill at
a beam rate of 10$^7$ particles per spill.
An online filter
performed a preliminary
track reconstruction in the M2 spectrometer. Requiring at least one
track with negative and none with positive slope together with other
conditions crucial to a complete reconstruction reduced this sample
by a factor of 1:1.7.

In the 1997 run \SELEX has recorded 215~million candidates for hadron-electron
scattering with the \PgSm/\Pgpm-beam.
In preparation for a first analysis with the software tools
available at that time the negative-beam sample was stripped to 10~\% of
its original size by cutting on an electron signature, unambiguous
identification of the beam particle, and a two-negative-track
event topology. A second-stage strip required a
two-prong vertex, again reducing the sample by a factor of 10.

Out of the stripped data sample described above, 12,000 \PgSm-~electron and
26,000 \Pgpm- ~electron elastic scattering events were extracted.
For each event, the incoming and outgoing tracks in the vertex
were required to be coplanar. Particle identification for the two outgoing
tracks was performed by combining information from he transition
radiation detectors with kinematic
constraints. Events with ambiguous particle identification were discarded.
\begin{figure}[h]
  \begin{center}
       \epsfig{file=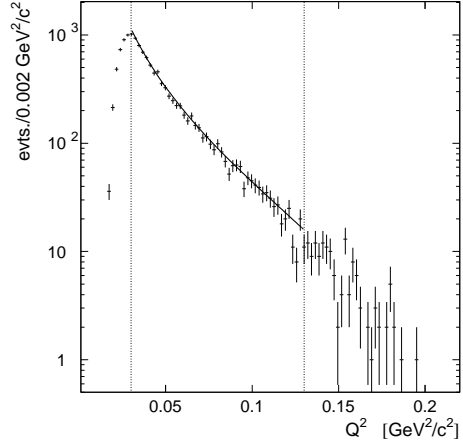, width=6.5cm}
      \caption{\small \Q2 distribution of \PgSm-electron scattering events.
        Vertical lines indicate the region accepted for fitting. }
      \label{fig:q2dist}
  \end{center}
\end{figure}
For \PgSm, decays upstream of
the M2 chambers were rejected by requiring the scattered beam particle to
have at least 60~\% of the incoming beam particle's momentum.
Finally, electron momentum and scattering angle had to match their expected
kinematic relation to better than 10~\%. 

The charge radii were determined by fitting the differential cross section
with an assumed radius as the single parameter to the observed distribution of
the four-momentum transfer squared \Q2 (Fig.~\ref{fig:q2dist}).
Since the shape of the \Q2-distribution yields the radius no absolute
normaliziation is needed.
%However, acceptance effects can have a significant impact on the
%result.
In this first analysis, \Q2 was calculated from the beam momentum and
the scattering angle of the electron. From Monte Carlo studies the
\Q2 resolution was estimated to be 1.5~\%. 
Preliminary acceptance studies were performed using generated elastic
scattering events
embedded in real data. The geometrical and reconstruction-dependent
acceptance was modeled and a preliminary evaluation of the trigger
efficiency performed.

\begin{figure}[h]
  \begin{center}
    \leavevmode
    \epsfig{file=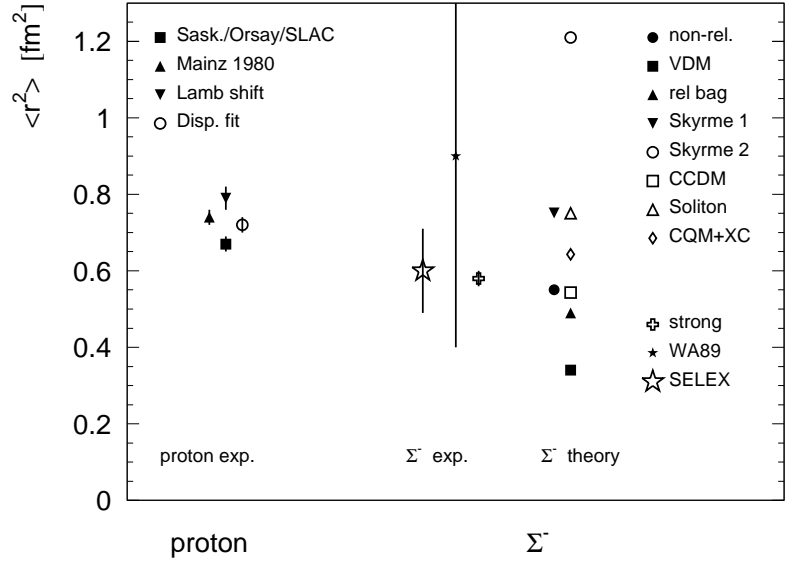, height=9cm}
  \end{center}
  \caption{The \PgSm\ charge radius compared to
    various results for the proton radius (left):
    \textit{Sask./Orsay/SLAC}\protect\cite{hand63,murphy74},
    \textit{Mainz}\protect\cite{simon80a},
    \textit{dispersion-theoretical fit} to all of
    above\protect\cite{mergell96}, and
    \textit{Lamb shift}\protect\cite{udem97}.
     -- Experimental results (center):
    \textit{SELEX}: this measurement,
    \textit{WA89}: WA89 result\protect\cite{wa89_radius_paper},
    \textit{strong}: strong interaction radius\protect\cite{povh90}. -- 
    The predictions for \PgSm\ refer to the following models:
    \textit{non-rel.}: non-relativistic
    quark model, \textit{VDM}: vector dominance model,
    \textit{rel bag}: relativistic bag model (all three values from
    \protect\cite{povh90}),
    \textit{Skyrme 1}: Skyrme model\protect\cite{kunz91},
    \textit{Skyrme 2}: Skyrme model\protect\cite{park92},
    \textit{CCDM}: Chiral color dielectric model\protect\cite{sahu94},
    \textit{Soliton}: Soliton model\protect\cite{kim96},
    \textit{CQM+XC}: Chiral constituent quark model including
    exchange currents\protect\cite{wagner98b}.}
  \label{fig:sigmaresult}
\end{figure}
For the \PgSm\ data, a \Q2 region with flat acceptance was chosen for
fitting the radius. For the \Pgpm\ data, an acceptance correction was applied.
Each event was normalized to its individual beam momentum to eliminate
effects of the beam momentum spread. An unbinned maximum likelihood
fit using dipole electric and magnetic form factors for the \PgSm\
yields a mean squared charge radius of
\begin{equation*}
  \label{equ:final_result}
  \r2_{\Sigma^-} = 0.60 \pm 0.08 \:(stat.) \pm 0.08 \:(syst.) \:\text{fm}^2
\end{equation*}
in the \Q2 region of
$  0.03 \le Q^2 \le 0.16 \; \text{GeV}^2/c^2 $ (7,800 events)\cite{ivo}.
This result is well inside
the limits determined by the WA89 collaboration \cite{wa89_radius_paper},
$0.4~\fm2 \le \r2_{\Sigma^-} \le 1.4~\fm2$ (Fig.~\ref{fig:sigmaresult}).

For the negative pion, a monopole electric form factor is used. We find
\begin{equation*}
  \label{equ:pion_selex}
  \r2_{\pi} = 0.45 \pm 0.03 \:(stat.) \pm 0.07 \:(syst.) \:\fm2,
\end{equation*}
where
$  0.03 \le Q^2 \le 0.20 \; \text{GeV}^2/c^2.$ (12,000 events)\cite{kvo_phd}.
This result is in excellent agreement with
the so far best direct measurement \cite{amendolia86pion} of
{\nobreak $\r2_{\pi} = 0.44 \pm 0.01 \:\fm2$} as well as a recent
calculation which takes into account form factor measurements in both
space-like and time-like regions \cite{geshkenbein98}:
{\nobreak $\r2_{\pi} = 0.463 \pm 0.005 \:\fm2.$}

Major contributions to the systematic error come from the \Q2 resolution,
uncertainties in the corrections for trigger efficiency, and beam
contamination by other particles, particularly \PgXm. Significant improvement
is expected for all of these when advanced reconstruction and simulation
software is used to refine the data sample. \Q2 will be determined from
all kinematic variables and events with identified \PgSm\ decays accepted
as well. We anticipate a statistical error
of less than 10~\% in the final analysis of the \PgSm\ radius.

%
% ---------------------- Sigma-p total X-section ------------------------
%

\section{\PgSm\Pp\ total cross section}
\label{sec:xsec}

In general, the difference of total hadronic cross sections 
is ascribed to the difference in Regge residue functions, which are connected
to hadronic radii, rather than to the Pomeron propagator.
In the Landshoff-Donnachie~\cite{donnachie92} version of Regge theory
the effective Pomeron intercept $\epsilon\approx(\alpha_P(0)-1)$ and the
effective Reggeon intercept $\eta\approx(\alpha_R(0)-1)$ are assumed to be
universal, i.e. the same for all hadrons, with $\epsilon \approx 0.08$ and
$\eta\approx 0.47$.

Recent data from H1 and ZEUS on the proton structure function at
small $x$ and high $Q^2$ demonstrate, however, that the effective Pomeron
intercept is higher for 
hadrons with smaller radii, up to $\epsilon = 0.4$ for high $Q^2$~\cite{hera}.
There is further evidence for $\epsilon > 0.08$ from real
exclusive photoproduction of heavy
flavors. Data from HERA also show that the cross section of $J/\Psi$
photoproduction rises by a factor of 6 from $\sqrt{s}$ = 10 to
100~GeV~\cite{j/psi,levy95}.
%Data on photoproduction of $\phi$ vector mesons also indicates faster
%growth \cite{landshoff95}.

This $Q^2$ dependence of the $\epsilon$ should have its counterpart in
the $\epsilon$ dependence on the radii of the stable hyperons. The higher
the quark mass, the smaller the interquark distance corresponding to the
effective high $Q^2$ hadronic interaction.

The only available beams with flavors heavier than $up$ or $down$ are
\PKpm\ and hyperon beams.
Total cross sections of $\Sigma^-$ and $\Xi^-$ on protons and deuterons
have been measured at beam energies between 19--137~GeV at
CERN~\cite{WA42,badier72}.
Analogous data at 600 GeV would provide a sensitive test of the Pomeron
universality \cite{boris}.

%Recent HERA data shows a much steeper increase with energy
%of the \PJgyy\ photoproduction cross section than it is observed for
%the light vector mesons. This effect is ascribed to the mass of the
%charm/anticharm quarks setting the scale for the interaction \cite{levy95}.
%A not as dramatic but similar effect may be seen on the slope of the
%$\Pgg\Pp\to\phi\Pp$ cross section. Unfortunately, currently available
%data does not allow a definitive answer. This effect may, however, be
%observed as well in the behavior of the
%total cross section $\sigma_{tot}(\PgSm\Pp)$ in comparison to
%$\sigma_{tot}(\Pp\Pp)$ at high energies.

In the 1997 fixed target run \SELEX has recorded 58 million events
with \PgSm/\Pgpm\ and 18 million with proton/\Pgpp\ beams using a
beam-only trigger.
Beam particles were identified with a transition radiation
detector. Corrections were applied to account for effects of Coulomb and
Coulomb-nuclear interference as well as beam rate and contamination.

\begin{figure}[h]
  \begin{center}
    \epsfig{file=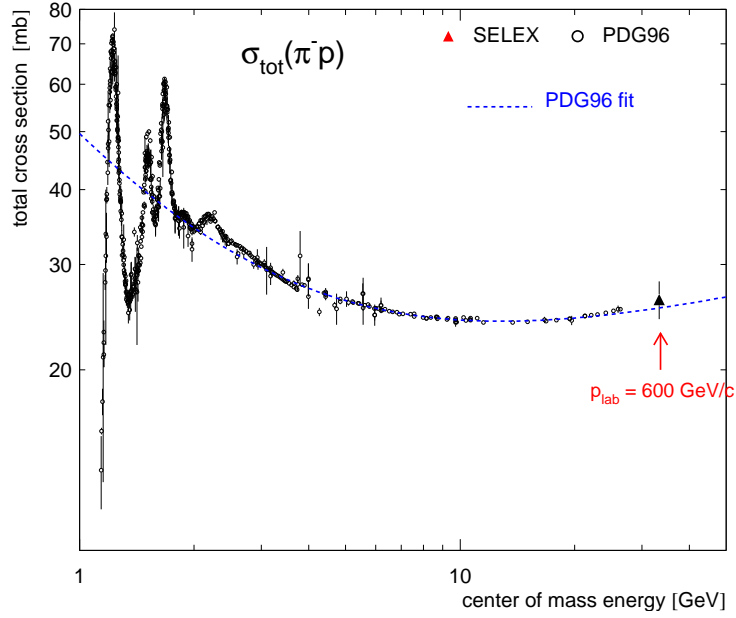, height=8.5cm}
    \caption{Compilation of world data on \Pgpm\Pp\ total cross sections
      with the preliminary \SELEX result at 600~\gevc beam momentum.}
    \label{fig:pi_xsec}
  \end{center}
\end{figure}
\begin{figure}[h]
  \begin{center}
    \epsfig{file=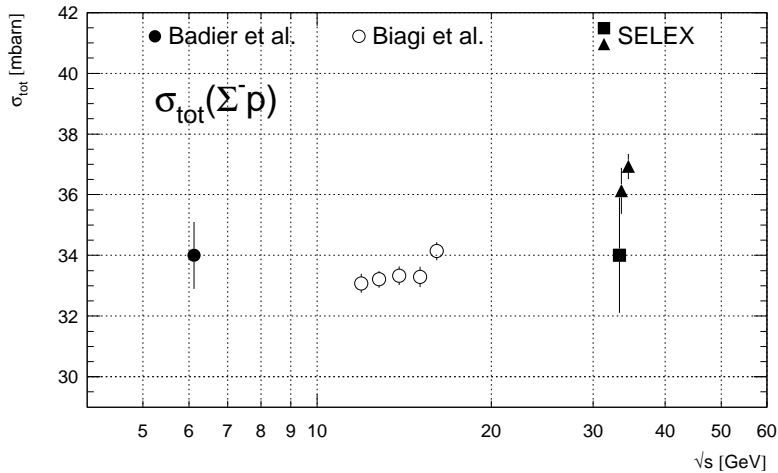, height=7cm}
    \caption{Total \PgSm\Pp\ cross section at different center-of-mass
      energies. Low-energy data from Badier et al.~\protect\cite{badier72}
      and WA42 (Biagi et al.~\protect\cite{WA42}).
      \SELEX results from C--CH$_2$ difference (square) and C and Be
      target Glauber calculations (triangles).}
    \label{fig:sigma_xsec}
  \end{center}
\end{figure}

The total hadronic \PgSm\Pp\ cross section at 600~\gevc beam momentum
has been determined  by the transmission method:
$${ \sigma_{tot}(\Omega)} := \frac{1}{\rho L}
\lim_{\Omega\rightarrow 0}
\log \left[
  \frac{F_{0}}{F_{tr}(\Omega)}\cdot
  \frac{E_{tr}(\Omega)}{E_{0}}\right ].
$$
Here, $\rho$ and $L$ are density and length of the target, and
$F_{tr}/F_{0}$ and $E_{tr}/E_{0}$
the transmission ratios with and without target, respectively.
Instead of a liquid hydrogen target \SELEX had two alternative approaches:
\begin{itemize}
\item[(1)] C--CH$_2$ subtraction method.
  From the data sample taken with carbon and polyethylene targets
  the total \PgSm-proton cross section was calculated from
  $$\sigma_{tot}(\Sigma^-p) = \frac{1}{2} \left[ 
    \sigma_{tot}(\Sigma^-CH_2) - \sigma_{tot}(\Sigma^-C)  \right].$$
  This yields
  $$\sigma_{tot}(\Sigma^-p) =  34.0 \pm 1.9 \:\text{mb},$$
  where statistical and systematic errors have been combined.
  As a cross check, an identical analysis was performed on the pion-proton
  data taken with these targets:
  $$\sigma_{tot}(\pi^-p) =  26.2 \pm 1.9 \:\text{mb}$$
  A comparison to the current world data sample which only covers beam
  momenta up to 370~\gevc \cite{pdg} finds our results to be following the
  general trend (Fig.~\ref{fig:pi_xsec}).
\item[(2)] $\sigma_{tot}(\PgSm\Pp)$ was calculated from the ratio of
  proton-nucleus and \PgSm-nucleus cross sections. Here, we used
  Be and C targets.
  The \PgSm-nucleus cross sections for Be and C targets follow nicely the
  expected $A$-dependence of $\sigma_{tot}(XA)=\sigma_0A^\alpha$, where
  $\alpha\simeq 0.77$ \cite{uwe}. The \PgSm-proton cross section was
  calculated from the ratio $\sigma_{tot}(\PgSm A)/\sigma_{tot}(\Pp A)$.
  Nuclear effects were accounted for by a model based on Glauber theory
  which included corrections for inelastic nuclear shadowing.
  This leads to the result of $\sigma_{tot}(\Sigma^-p) = 36.39 \pm 0.76$~mb
  (Be target, 635~\gevc average beam momentum) and
  $\sigma_{tot}(\Sigma^-p) = 36.13 \pm 0.42$~mb (C target, 595~\gevc).
\end{itemize}
The results are shown along with total \PgSm\Pp\ cross sections at
lower energies in Fig.~\ref{fig:sigma_xsec}.
A fit to these data points using the
Donnachie-Landshoff parametrization \cite{pdg} yields
$\epsilon = 0.098 \pm 0.019$~\cite{uwe}.

%
% ------------------------ S1385 radiative width ---------------------------
%

\begin{figure}[hp]
  \begin{center}
    \epsfig{file=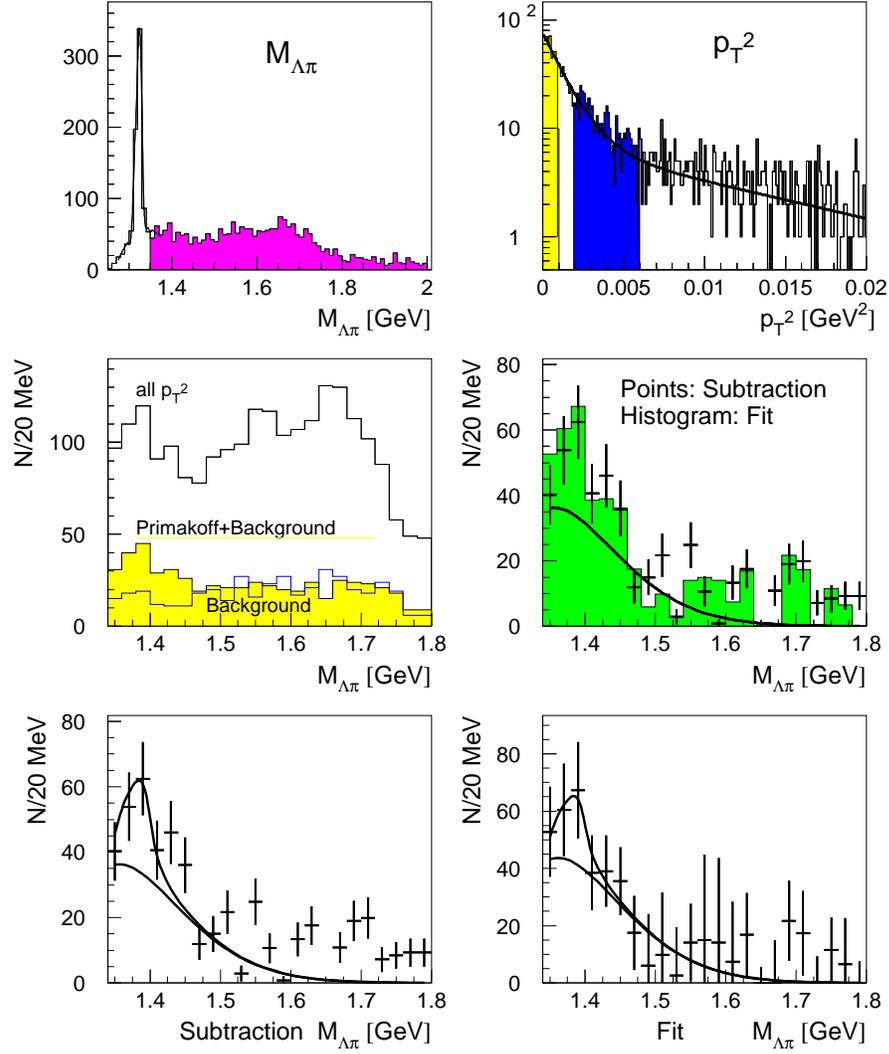, width=\textwidth}
    \caption{Transverse momentum squared and
      mass spectrum of the final state of
      $\PgSm + \text{Pb}\rightarrow \text{Pb}+ (\Pgpm + \PgL),
      \: \PgL \rightarrow \Pp + \Pgpm$. Explanations in text
      (section \ref{sec:1385}).}
    \label{fig:1385}
  \end{center}
\end{figure}

\section{Radiative width of \S1385}
\label{sec:1385}

Radiative decay widths of hyperons constitute powerful tests for
dynamical theories of hadronic systems. The expected value of the
SU(3)-suppressed radiative width $\Gamma(\Sigma^{*-} \to \PgSm\Pgg)$ in
different models is predicted in the region of 1--10~keV and the
SU(3)-allowed width $\Gamma(\Sigma^{*+} \to \PgSp\Pgg)$ in the
range of 100--300~keV \cite{width_theo}.
Unfortunately, the experimental situation is difficult due to small
branching ratios on one hand and large background from hadronic decays
on the other.

%Unfortunately, the branching
%ratios are very small. In addition, large background
%from strong decays with a \Pgpz\ in the final state where one of the
%decay photons has been lost makes it difficult to measure radiative
%widths directly.

The production of a hadron resonance state in the nuclear Coulomb field
(the Primakoff formalism), on the other hand, provides a relatively clean
method for the determination of radiative widths.

At \SELEX, the \S1385\ was produced from \PgSm\ using a lead target.
The differential cross section for the Primakoff reaction
$$ \PgSm + Z \to Z + \S1385,\:\: \S1385 \to \PgL + \Pgpm$$
can be written as a function of the four-momentum transfer squared $t$,
$$ \frac{d\sigma}{dt}=8\pi\alpha Z^2\:\frac{2J_{\Sigma^*}+1}{2J_{\Sigma}+1}
\:\Gamma(\S1385\to\PgSm\Pgg)\:\biggl(
\frac{m_{\Sigma^*}}{m_{\Sigma^*}^2-m_{\Sigma}^2}\biggr)^3\:
\frac{t-t_{min}}{t^2}\bigl|F(t)\bigr|^2,$$
%$$\frac{d\sigma}{dt dM^2}=\frac{Z^2\alpha}{\pi} 
%\frac{\sigma_{tot}}{M^2-m_{\Sigma^-}^2}
%\frac{t-t_{min}}{t^2} |F(t)|^2,$$
where $\alpha$ is the fine structure constant, $t_{min}$ the minimal
momentum transfer squared, and $m_{\Sigma^*}$ the mass of the final state.
$Z$ is the charge  and $F(t)$ the electromagnetic form factor of the nucleus.
The $t$-distribution for the Primakoff reaction has a pronounced forward
peak at $t=2 t_{min}$.

%The data was taken with a lead target.
The beam \PgSm\ was identified with a TRD. Elasticity of the reaction
and less than 2~GeV
of energy deposition in the first lead glass calorimeter were required.
The observed \PgL\Pgpm\ mass distribution
is shown in Fig.~\ref{fig:1385} (upper left), with the peak from \PgXm\
decays in the lead target clearly visible. 
The observed $p_T^2$ distribution (Fig.~\ref{fig:1385}, upper right)
was assumed to be the sum of
coherent \PgL\Pgp\ and Primakoff production.
From Monte Carlo simulation we found that both
coherent and Primakoff $p_T^2$ distributions smeared by the experimental
resolution can be described by double-exponential fits \cite{h-809}.

Two methods were used to estimate the \PgL\Pgp\ mass distribution for the
Primakoff reaction:
\begin{enumerate}
\item Two $p_T^2$ regions were defined (Fig.~\ref{fig:1385}, upper right),
  one at $p_T^2 < 0.001$~\gev2c2 where the Primakoff effect dominates,
  and one at $0.002 < p_T^2 < 0.006$~\gev2c2 for estimation of the
  background from coherent production of the \PgL\Pgp\ system. The
  corresponding mass spectra are shown in Fig.~\ref{fig:1385}, center left.
  Subtraction yields the data points in Fig.~\ref{fig:1385}, center right
  and lower left.
\item All events were subdivided into 20~MeV bins of the mass spectrum of
  the final \PgL\Pgp\ state, and the $p_T^2$ distribution analyzed for
  each bin. The result is shown as the shaded histogram in Fig.~\ref{fig:1385},
  center right and lower right.
\end{enumerate}
Both methods are in reasonable agreement (Fig.~\ref{fig:1385}, center right).
The total cross section  for the process
is given by the equation
\begin{equation*}
  \sigma_{tot}=\int_0^{\infty} \frac{d\sigma}{dt}dt=
  A\cdot\Gamma(\S1385 \to \PgSm\Pgg).
\end{equation*}
$A$ was obtained by integrating numerically over the differential cross section
$d\sigma/dt$.
The radiative width $\Gamma(\S1385 \to \PgSm\Pgg)$ was estimated using
the expression
\begin{equation*}
  \Gamma(\S1385 \to \PgSm\Pgg) = \frac{N_{\Sigma^*}}{A\cdot L\cdot\epsilon
   \cdot BR(\Sigma^*\to \PgL\Pgp)\cdot BR(\PgL \to \Pp\Pgp)},
\end{equation*}
where $N_{\Sigma^*}$ is the number of observed events, $L$ the luminosity
of the experiment, $\epsilon$ the combined reconstruction efficiency
where efficiency of the applied cuts and finite decay volume have been
accounted for. The luminosity was determined on the basis of coherent
production of $(\Pgpm\Pgpm\Pgpp)$ by pions \cite{zielinski83}.

The upper limit for the Primakoff production of \S1385 is: $N_{\Sigma^*}<205$
events at 95~\% confidence limit. With the above equation this yields
\begin{equation*}
  \Gamma(\S1385 \to \PgSm\Pgg) < 12\: \text{keV}\:(95~\%\:\text{CL}),
\end{equation*}
thus improving the upper limit of 24~keV established by an experiment
at Brookhaven (1977)~\cite{arik77}.

%
% ------------------------ conclusion ---------------------------
%

\section{Conclusions}
\label{sec:concl}

A measurement of the \PgSm\ mean squared charge radius has
been performed by elastic \PgSm-electron scattering.
A preliminary analysis yields a \PgSm\ radius of \break
{\nobreak
$\r2_{\Sigma^-} = 0.60 \pm 0.08 \:(stat.) \pm 0.08 \:(syst.) \:\text{fm}^2$}.
The \Pgpm\ radius was determined in parallel and is found to be
in excellent agreement with previous experiments.

The \PgSm\Pp\ total hadronic cross section has been measured at 600~\gevc.
We obtain {\nobreak $\sigma_{tot}(\PgSm\Pp)\:=\: 34.0 \pm 1.9$~mb}
from the difference of results for CH$_2$ and C targets, and
{\nobreak $\sigma_{tot}(\PgSm\Pp)\:=\: 36.6 \pm 0.9$~mb} from a Glauber model
calculation using the ratios of $\PgSm A$ to $\Pp A$ cross sections on
C and Be targets.

A new upper limit for the radiative width of the \S1385 has been established
at 12 keV (95~\% CL) from a study of Primakoff production on lead nuclei.

Improved statistics and smaller systematic errors for these results as well
as other hyperon physics results are expected as the analysis of \SELEX
data proceeds.

\section*{Acknowlegdements}
We are indebted to B.~C.~LaVoy, D.~Northacker, F.~Pearsall, and J.~Zimmer
for invaluable technical support.
This project was supported in part by
Bundesministerium f\"ur Bildung, Wissenschaft, Forschung und Technologie,
Consejo Nacional de Ciencia y Tecnolog\'{\i}a {\nobreak (CONACyT)},
Conselho Nacional de Desenvolvimento Cient\'{\i}fico e Tecnol\'ogico,
Fondo de Apoyo a la Investigaci\'on (UASLP),
Funda\c{c}\~ao de Amparo \`a Pesquisa do Estado de S\~ao Paulo (FAPESP),
the Israel Science Foundation founded by the Israel Academy of
  Sciences and Humanities,
Istituto Nazionale de Fisica Nucleare (INFN),
the International Science Foundation (ISF),
the National Science Foundation (Phy \#9602178),
NATO (grant CR6.941058-1360/94),
the Russian Academy of Science,
the Russian Ministry of Science and Technology,
the Turkish Scientific and Technological Research Board (T\"{U}B\.ITAK),
the U.S. Department of Energy (DOE grant DE-FG02-91ER40664),
and
the U.S.-Israel Binational Science Foundation (BSF).

%
%  --------------------- bibliography ----------------------
%

\end{document}